\documentclass[aps,pra,twocolumn,superscriptaddress,letterpaper]{revtex4-2}
\usepackage{svg}
\usepackage{chngcntr}
\usepackage{graphicx}
\usepackage{amsmath}
\usepackage{amssymb}
\usepackage{esint}
\usepackage{verbatim}
\usepackage{xcolor}
\usepackage{soul}
\usepackage{siunitx}
\usepackage{ulem}
\usepackage[utf8]{inputenc}
\usepackage[T1]{fontenc}
\usepackage[hidelinks]{hyperref}
\hypersetup{
    colorlinks=true,
    linkcolor=blue,
    filecolor=blue,      
    urlcolor=blue,
    citecolor=blue,
}

\newcommand{\be}{\begin{equation}}
\newcommand{\ee}{\end{equation}}
\newcommand{\ba}{\begin{array}}
\newcommand{\ea}{\end{array}}
\newcommand{\bqa}{\begin{eqnarray}}
\newcommand{\eqa}{\end{eqnarray}}

\usepackage{siunitx}

\newcommand{\nbathp}{n_{\text{p}}}

\newcommand{\Tf}{T_{\text{f}}}
\newcommand{\nfridge}{n_{\text{f}}}

\newcommand{\gOM}{g_{\text{OM}}}
\newcommand{\nth}{n_{\text{th}}}
\newcommand{\etaOM}{\eta_{\text{OM}}}

\newcommand{\ncav}{n_{\text{c}}}
\newcommand{\nwg}{n_{\text{wg}}}

\newcommand{\kappae}{\kappa_{\text{e}}}
\newcommand{\gammap}{\gamma_{\text{p}}}

\newcommand{\gammaOM}{\gamma_{\text{OM}}}

\newcommand{\Pin}{P_{\text{in}}}

\newcommand{\gammaSB}{\Gamma_\text{SB,0}}
\newcommand{\gammaPump}{\Gamma_\text{pump}}
\newcommand{\gammaDark}{\Gamma_\text{dark}}
\newcommand{\etaSPD}{\eta_\text{SPD}}

\begin{document}

\title{High-Efficiency Low-Noise Optomechanical Crystal Photon-Phonon Transducers}
\author{Sameer Sonar}
\thanks{these authors contributed equally to this work}
\author{Utku Hatipoglu}
\thanks{these authors contributed equally to this work}
\author{Srujan Meesala}
\author{David Lake}
\author{Hengjiang Ren}
\thanks{current address: Anyon Computing Inc., Emeryville, CA 94608}
\affiliation{Institute for Quantum Information and Matter and Thomas J. Watson, Sr., Laboratory of Applied Physics, California Institute of Technology, Pasadena, California 91125, USA}
\affiliation{Kavli Nanoscience Institute, California Institute of Technology, Pasadena, California 91125, USA}

\author{Oskar Painter}
\thanks{opainter@caltech.edu; http://copilot.caltech.edu}
\affiliation{Institute for Quantum Information and Matter and Thomas J. Watson, Sr., Laboratory of Applied Physics, California Institute of Technology, Pasadena, California 91125, USA}
\affiliation{Kavli Nanoscience Institute, California Institute of Technology, Pasadena, California 91125, USA}
\affiliation{ AWS Center for Quantum Computing, Pasadena, CA,
USA}

\date{\today}

\begin{abstract} 

Optomechanical crystals (OMCs) enable coherent interactions between optical photons and microwave acoustic phonons, and represent a platform for implementing quantum transduction between microwave and optical signals. Optical absorption-induced thermal noise at cryogenic (millikelvin) temperatures is one of the primary limitations of performance for OMC-based quantum transducers. Here, we address this challenge with a two-dimensional silicon OMC resonator that is side-coupled to a mechanically detached optical waveguide, realizing a six-fold reduction in the heating rate of the acoustic resonator compared to prior state-of-the-art, while operating in a regime of high optomechanical-backaction and millikelvin base temperature. This reduced heating translates into a demonstrated phonon-to-photon conversion efficiency of {93.1 $\pm$ 0.8}\% at an added noise of 0.25 $\pm$ 0.01 quanta, representing a significant advance toward quantum-limited microwave-optical frequency conversion and optically-controlled quantum acoustic memories.


\end{abstract}

\maketitle


Optomechanical crystals (OMCs) are periodic dielectric structures engineered to co-localize light and acoustic vibrations on the wavelength scale \cite{Eichenfield2007}. OMCs in one-dimensional silicon nanobeams have been used to demonstrate MHz-scale interaction rates between single photons and phonons confined in telecom-wavelength optical and GHz-frequency acoustic modes, respectively \cite{Chan2012}. In recent years, these devices have enabled advances in quantum acoustics \cite{maccabe2020nano, wallucks2020quantum}, and microwave-optical quantum transduction \cite{mirhosseini2020superconducting, meesala2024non, jiang2023optically, weaver2023integrated}. In these applications, OMCs are operated at temperatures $\leq0.1$ K to ensure negligible thermal occupation in GHz frequency acoustic modes. However, optical excitation of OMCs in this temperature regime is accompanied by local heating due to weak parasitic absorption of laser light. This effect results in elevated thermal occupation and reduced coherence of the acoustic mode \cite{meenehan2015pulsed,maccabe2020nano}, and limits the performance of OMCs as quantum memories and transducers. For instance, in the context of microwave-optical quantum transduction, operation of OMCs in the quantum coherent regime is currently possible in pulsed mode with reduced laser power, albeit at the expense of reduced efficiency for the optomechanical scattering process, and ultimately, a reduced entanglement generation rate for quantum networking applications \cite{mirhosseini2020superconducting,meesala2024non}.

Two-dimensional (2D) device geometries \cite{safavi2014two,ren2020two,madiot2023multimode,povey2024two} which offer increased thermal contact with the substrate are a promising approach to mitigate the detrimental effects of optical absorption-induced heating in OMCs at millikelvin temperatures. These structures rely on a phononic bandgap to protect the acoustic mode of interest from clamping losses while leveraging a larger density of states at frequencies above the gap to allow high frequency thermal phonons to escape out of the OMC cavity. However, a large thermal contact area can also lead to a significant influx of thermal phonons from sites of optical absorption external to the cavity. Such a phenomenon was noted in our prior work on 2D OMCs in Ref.\,\cite{ren2020two}, where optical absorption in the coupling waveguide was found to be the dominant source of thermal noise in the acoustic mode. In principle, thermal isolation of the OMC cavity from the coupling optical waveguide could be achieved through mechanical detachment of both structures from one another while maintaining evanescent optical coupling. However, these requirements are challenging to engineer in a 2D geometry since abrupt termination of the OMC can result in significantly increased optical loss via parasitic edge modes. In this work, we overcome this challenge with a 2D OMC with a novel side-coupled waveguide design and demonstrate a significant improvement in phonon-to-photon conversion performance over previous state-of-the-art \cite{ren2020two}. We observe a six-fold reduction in the heating rate of the acoustic resonator in the regime of high optomechanical backaction and millikelvin base temperatures. Under continuous wave (pulsed) optical excitation, we show through optomechanical sideband thermometry that the device can perform phonon-to-photon conversion with an internal efficiency of 98.6 $\pm$ 0.2\% (93.1 $\pm$ 0.8\%) and an added noise of 0.28 $\pm$ 0.01 quanta (0.25 $\pm$ 0.01 quanta) at the highest optical powers used in our experiments. Our results indicate that side-coupled 2D OMCs can enable high-fidelity quantum-coherent operations with acoustic modes and advance the performance of optomechanical quantum transducers and memories.

\begin{figure*}[tp]
\begin{center}
\includegraphics[width=2\columnwidth]{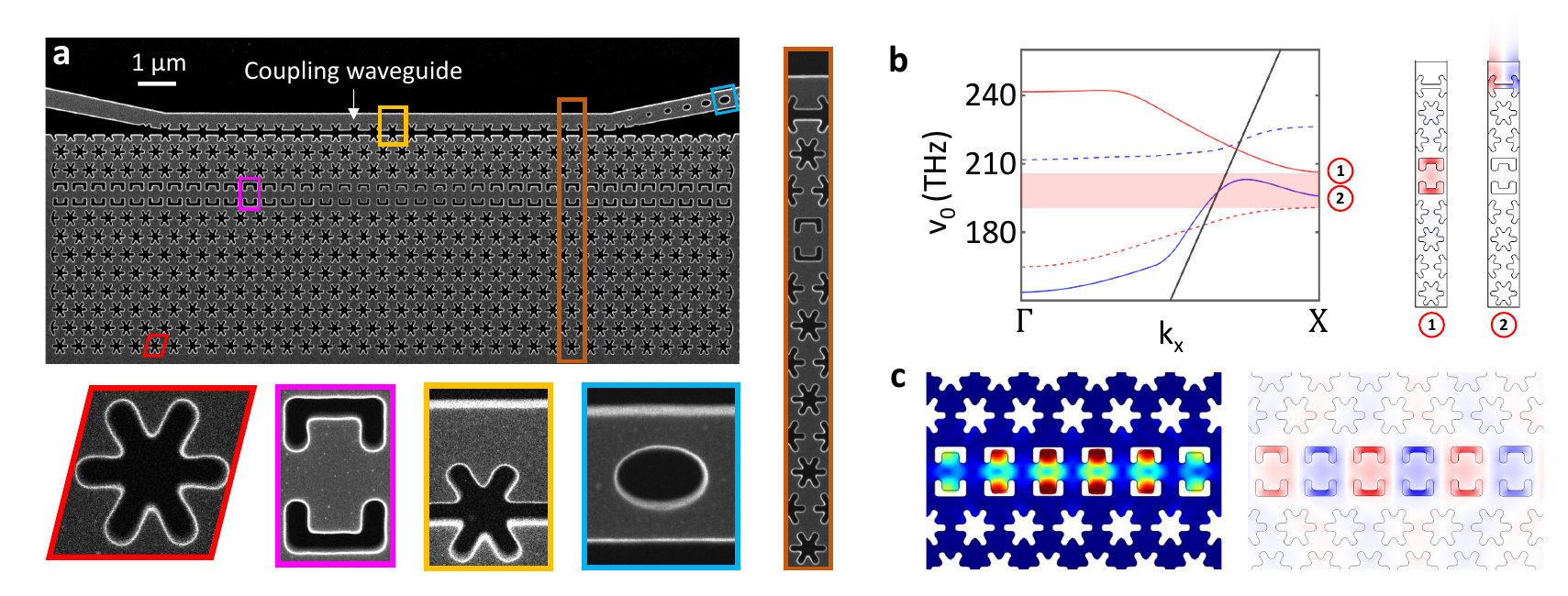}
\caption{\textbf{Side-coupled 2D optomechanical crystal (OMC) cavity. a,} Helium-ion microscope image of a representative device with insets indicating salient features from left to right: unit cells of the 2D snowflake lattice, central fish-bone waveguide, optical coupling waveguide, and the optical waveguide mirror. The orange inset on the right shows the supercell of the geometry used to simulate optical and acoustic bandstructures. \textbf{b}, Simulated optical bandstructure of the supercell. The solid red (blue) band has energy predominantly in the cavity (coupling waveguide). Transverse electric field profiles of these optical modes at the X-point are shown on the right. Dashed lines in the bandstructure indicate other guided modes. \textbf{c}, FEM simulations of the acoustic (left; total displacement) and optical (right; transverse electric field) modes of the 2D OMC cavity with acoustic resonance frequency, $\Omega_\text{m}/2\pi=10.3$ GHz and optical resonance wavelength, $\lambda=1550$ nm, respectively. 
}
\label{fig1}
\end{center}
\end{figure*}


{}

\vspace{2mm}
\noindent\textbf{Design of the side-coupled 2D-OMC cavity.} 
\noindent The OMC cavity in this work is designed on a 220 nm thick silicon device layer of a silicon-on-insulator (SOI) substrate. Figure \ref{fig1}(a) shows a helium ion microscope image of a fabricated device. The insets show (from left to right) unitcells of snowflake, fish-bone waveguide, coupling waveguide, and the photonic crystal mirror. The coupling waveguide is designed with a half-snowflake unitcell to evanescently couple to the optical mode of the OMC cavity. The gap between the two structures can be controlled to set the external coupling rate of the optical cavity. To facilitate measurements of the OMC in reflection, the coupling waveguide is terminated with a photonic crystal mirror. The orange inset on the right shows the supercell for the combined setup with the cavity and coupling waveguide. The corresponding optical bandstructure is shown in Figure \ref{fig1}(b). The shaded region represents the relevant optical bandgap. The cavity (waveguide) band  of interest is shown in solid red (blue). The corresponding electric field profile for cavity (waveguide) mode is shown on the right labeled 1 (2).


Figure \ref{fig1}(c) illustrates the simulated acoustic breathing mode with a frequency of 10.3 GHz, and the transverse electric field distribution of the fundamental optical mode with a wavelength of 1550 nm. The energy of the optical mode is predominantly localized in the air gaps of the fish-bone structure to enhance optomechanical coupling due to the moving boundaries. For an air gap size of 70 nm, this design provides a vacuum optomechanical coupling rate, $\gOM/2\pi=$ 1.1 MHz in simulation. In principle, fabrication of smaller gaps (e.g. 20 nm) can allow for $\gOM/2\pi$ up to 2.5 MHz. More details are provided in supplementary text, section VII. 

The devices are patterned using electron beam lithography and reactive ion etching, and are suspended by etching the underlying buried oxide layer with hydrofluoric acid etch. We characterized two devices in a dilution refrigerator at a base temperature of $\Tf\approx$ 10 mK. The optical and acoustic mode parameters of both devices are tabulated in supplementary text,
section I.

\begin{figure*}[tp]
\begin{center}
\includegraphics[width=2\columnwidth]{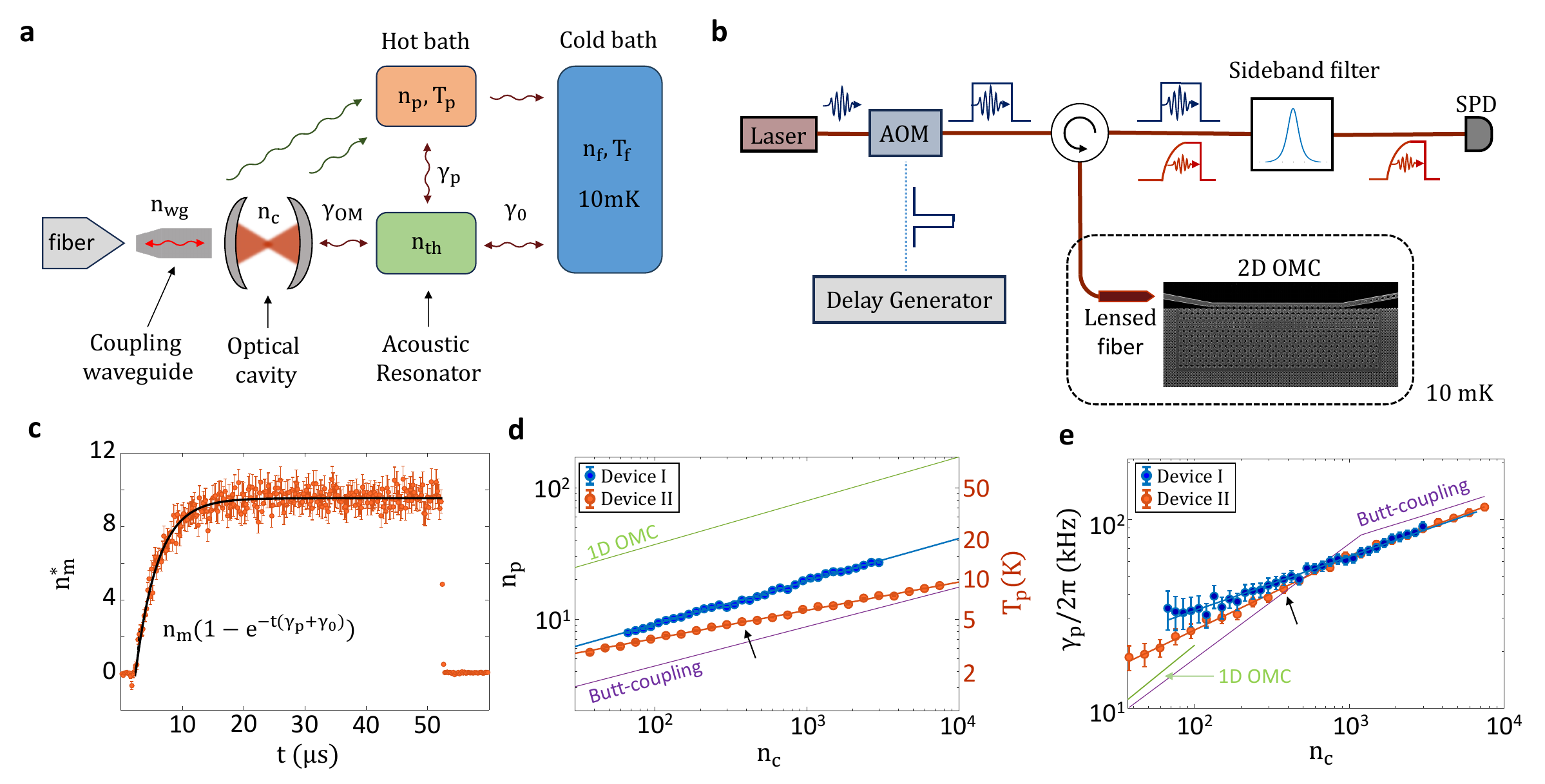}
\caption{\textbf{Characterization of optical absorption-induced hot bath. a,} Schematic showing interactions of the acoustic resonator with various baths considered in our heating model. \textbf{b,} {Schematic of measurement setup for time-resolved measurements of the hot bath using single-photon counting on the optical sideband generated by thermal motion of the acoustic resonator.} \textbf{c,} Measurement of the transient thermal occupation of the acoustic resonator, $n_\text{m}^\star$ in response to a rectangular optical pulses on resonance with the optical cavity (pulse duration $\tau_\text{d}= \SI{50}{\micro\second}$, repetition rate, $R$ = 1 kHz, and peak optical photon occupation, $\ncav$ = 385). The black line represents an exponential fit to the observed data with the characteristic rate, $\gammap+\gamma_0$, and steady-state thermal occupation, $n_\text{m}$. Here $\gamma_0$ is the intrinsic damping rate of the acoustic resonator and $\gammap$ is the coupling rate to the optical-absorption-induced hot bath. \textbf{d,} Thermal occupation of the hot bath, $\nbathp$ estimated from measurements of $n_\text{m}$ performed at varying optical power, shown on the x-axis in units of peak intra-cavity photon occupation, $\ncav$. For comparison, $n_p$ curves for 1D-OMC \cite{maccabe2020nano}, butt-coupled 2D OMC \cite{ren2020two} are shown. \textbf{e,} Variation of $\gammap/2\pi$ with $\ncav$. The data point marked with an arrow in panels (d) and (e) corresponds to the data in panel (c) for $\ncav=385$.}
\label{fig2}
\end{center}
\end{figure*}

\vspace{2mm}
\noindent\textbf{Optical absorption-induced hot bath.} Previous measurements on OMC devices in the dilution refrigerator have shown that the acoustic mode thermalizes to temperatures well below 100 mK \cite{maccabe2020nano}. In our study, we model this connection to the cold substrate by a coupling to a cold bath with occupancy \(n_\text{f}\) ($<10^{-3}$), at an intrinsic acoustic damping rate, $\gamma_0$, as shown in Figure \ref{fig2}(a). Under excitation with laser fields, optical absorption-induced heating is modeled by considering a hot bath at a thermal occupation, $n_\text{p}$ (corresponding to a bath temperature, $T_\text{p}$), coupled to the acoustic mode at a rate $\gamma_\text{p}$. The acoustic resonator experiences optomechanical back-action at a rate, $\gamma_\text{OM}$. When the laser is tuned to the red motional sideband of the optical cavity  (detuning $\Delta=-\Omega_\text{m}$), a parametric beamsplitter interaction allows us to operate the device as a linear, bi-directional converter between quantum states in the acoustic and optical modes at a rate, $\gamma_\text{OM}=4g_\text{OM} n_\text{c}/\kappa_\text{t}$ where $\kappa_\text{t}$ is the total linewidth of the optical resonance \cite{Safavi-Naeini2011a}.
However, as a first step, we characterize the optical absorption-induced hot bath in our devices by operating the laser on resonance with the optical cavity ($\Delta=0$). We measure the heating dynamics of the acoustic mode by detecting optical photons scattered from the laser field onto a motional sideband of the optomechanical cavity. In steady state, the occupation of the acoustic mode, $n_\text{m}$ is expected to be an average of the thermal occupations of the hot and the cold baths, weighted by the coupling rates of the acoustic mode to the respective baths, as given by the relation,
\begin{equation}
    n_\text{m}=\frac{\gamma_\text{p} n_\text{p}  + \gamma_0 n_\text{f}}{\gamma_\text{p} + \gamma_0 }.
    \label{eq:nm}
\end{equation}

\begin{figure*}[tp!]
\includegraphics[width=\textwidth]{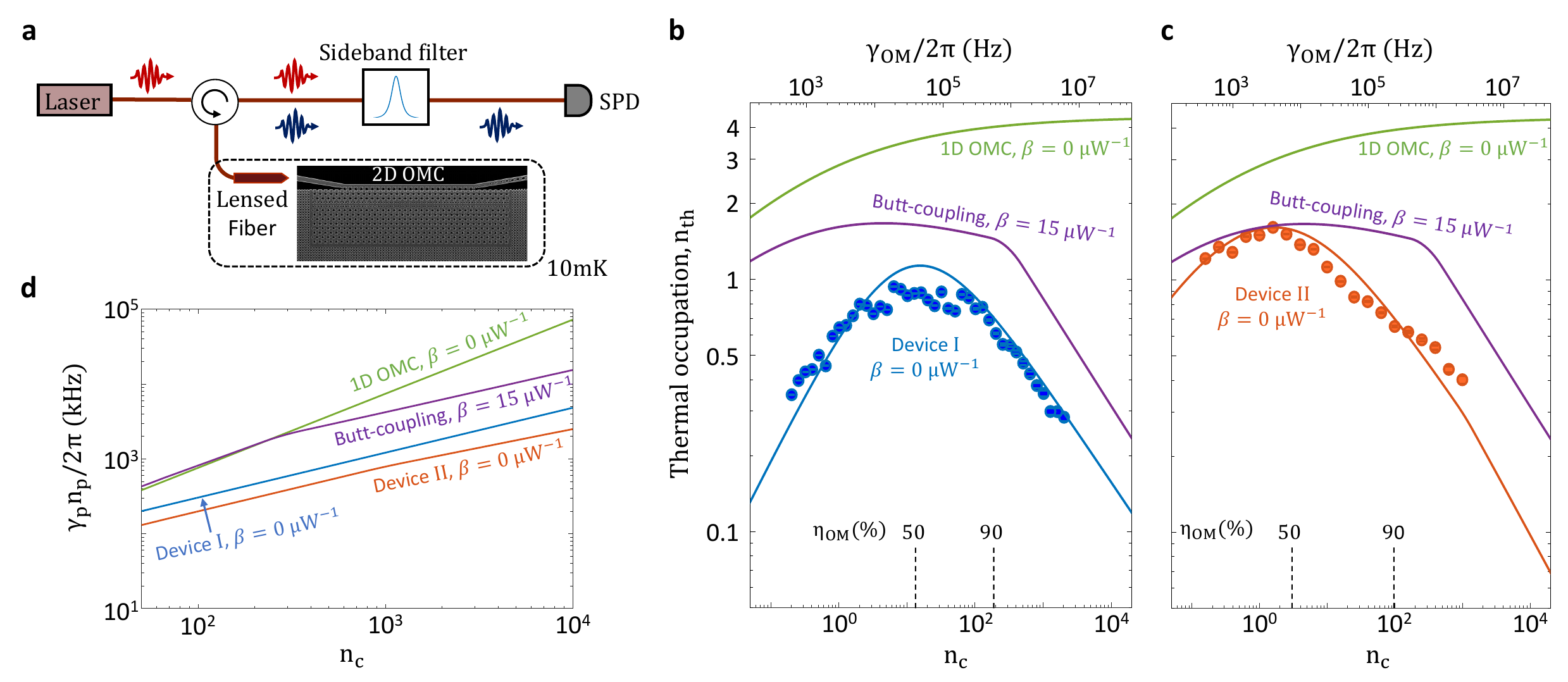}
\caption{\textbf{Phonon-to-photon transduction under continuous-wave excitation. a,} Schematic of measurement setup showing single-photon counting of up-converted photons at the optical resonance frequency with the OMC pumped continuously on the red-detuned sideband ($\Delta=-\Omega_\text{m}$) of the optical resonance. Measured thermal phonon occupancy, $\nth$ with varying optical power, shown on the bottom x-axis in units of intra-cavity photon occupation, $n_\text{c}$, and on the top-axis in units of optomechanical transduction rate, $\gammaOM$. Results are shown on separate charts for \textbf{b} device I, and \textbf{c,} device II. Filled circles are data points whereas the solid line indicated with $\beta=\SI{0}{\micro\watt}^{-1}$ is the modeled $\nth$ dependence using equation \ref{eq:nth_theory}. For comparison, model curves are shown for a butt-coupled 2D-OMC ($\beta=\SI{15}{\micro\watt}^{-1}$) \cite{ren2020two}, and 1D-OMC \cite{maccabe2020nano}. Dashed lines indicate the $n_\text{c}$ value for optomechanical transduction efficiency $\etaOM= 50\%$ and $90\%$. For $n_\text{c}=1$, on chip input power for device I and II are $\Pin= \SI{0.20}{\micro\watt}$, and $\SI{0.28}{\micro\watt}$, respectively. \textbf{d,} {Estimated heating rate of the acoustic resonator $\gamma_\text{p} n_\text{p}/2\pi$ as a function of $n_\text{c}$ under $\Delta=-\Omega_\text{m}$ for different devices plotted for their measured value of $\beta$.}}
\label{fig3}
\end{figure*}

Figure \ref{fig2}(b) shows the schematic of the measurement setup for $n_\text{m}$. We send laser pulses with pulse duration, $\tau_\text{d} = \SI{50}{\micro\second}$ at a repetition rate, $R$ = 1 kHz, to the device in the DR via a circulator. The optical signal reflected from the device is directed to a Fabry-Perot filter setup which suppresses the pump pulses and transmits photons generated on the motional sideband of the optomechanical cavity to a superconducting nanowire single photon detector. Figure \ref{fig2}(c) shows the time-dependent occupation of the acoustic mode measured from heating induced by a square laser pulse with a peak power corresponding to $n_\text{c}=385$. The rate of increase in the occupation is used to infer $\gamma_\text{p} + \gamma_0$, whereas the steady state occupation is used to infer $n_\text{m}$. Finally, the value of $\gamma_0$ is measured independently from ringdown measurements \cite{maccabe2020nano}, thereby allowing us to extract the parameters, $\gamma_\text{p}$ and $n_\text{p}$ of the hot bath from Equation \ref{eq:nm}. 

Figure \ref{fig2}(d) shows $n_\text{p}$ as a function of $n_\text{c}$ for devices I and II along with results from similar measurements performed previously on 1D OMCs \cite{maccabe2020nano}, and butt-coupled 2D OMCs \cite{ren2020two}. The vertical axis on the right represents the corresponding bath temperature $T_\text{p}$. The solid lines around the experimental data points for devices I and II are fits to the power-law, $\nbathp=A n_\text{c}^{k}$. We find that the power law exponents for device I and II are 0.31 and 0.21, respectively. In comparison, the power laws for $n_\text{p}$ on butt-coupled 2D OMC and 1D OMC cavities were found to have exponents of 0.3 and 0.33, respectively. The variations in \(n_\text{p}\) power laws for side-coupled devices deviate from predictions made by the phonon-bottleneck model \cite{maccabe2020nano}, suggesting underlying mechanisms that require systematic study beyond the scope of this article. However, we note that previous studies on phononic crystal structures have observed varying power laws for thermal conductance \cite{zen2014engineering}, and disorder dependent thermal conductance \cite{ maire2017heat,wagner2016two}, which may offer insights into the observed $n_\text{p}$ power laws in this work.

For a given optical power, the magnitude of $n_\text{p}$ for side-coupled 2D OMCs in this study is higher compared to that on a butt-coupling 2D OMCs. This relative increase in bath temperature can be attributed to the cut in the silicon device layer required for the side-coupled waveguide, which reduces the in-plane solid angle over which the cavity is connected to the cold substrate in comparison with the butt-coupled geometry. Nonetheless, we will show in the following measurements that the side-coupled design ultimately provides significantly lower heating rate in the practically relevant $\Delta=-\Omega_\text{m}$ setting used for optomechanical transduction.

Figure \ref{fig2}(e) shows the variation of $\gammap$ with $\ncav$. For device II, we find the power-law, $\gammap/2\pi= 4.3$ (kHz)$\times \ncav^{0.39}$ when $\ncav < 1000$ and $\gammap/2\pi= 8.25$ (kHz)$\times \ncav^{0.29}$ when $\ncav > 1000$. The power-law exponent of 0.39 in the regime of low optical power is significantly different from 0.6 and 0.66 observed previously for butt-coupled 2D OMC and 1D-OMC devices, respectively. 
{This reduction can be explained by the reduced $\nbathp$ power exponent of 0.21, together with a two-dimensional phonon bath (see supplementary text, section X).}
For device I, in the regime of high optical power with $\ncav > 1000$, $\gammap/2\pi= 8.7$ (kHz)$\times \ncav^{0.29}$. For $\ncav < 1000$, the measurement error on $\gammap$ for device I is larger due to a higher intrinsic damping rate, $\gamma_0$, thereby reducing the reliability of a fit in this regime. We observe that the power-law exponent for $\gammap$ in the regime of high optical power is identical for devices I and II in this study, and also in close agreement with the butt-coupling geometry. 


\vspace{2mm}

\noindent \textbf{Phonon-to-photon transduction under continuous-wave laser excitation.} 
After characterizing the optical-absorption-induced hot bath, we test the device with the laser tuned to the red motional sideband of the optical cavity ($\Delta=-\Omega_\text{m}$), relevant for phonon-photon transduction. The thermal occupation of the acoustic mode, $\nth$, due to optical-absorption-induced heating adds finite noise to any transduced signal. Other performance metrics of interest for such a transducer are the conversion efficiency, $\etaOM$, and bandwidth, $\gamma_\text{m}$. 
In continuous-wave operation, the conversion efficiency is given by $\etaOM=\frac{\gammaOM}{\gammaOM+\gamma_0+\gammap}$. We calibrate $\gammaOM$ for different $\ncav$ using electromagnetically induced transparency  \cite{safavi2011electromagnetically} (see supplementary text, Figure S3(a) for details). Due to the low $\gamma_0$ in our devices (21 kHz for device I and 0.97 kHz for device II), we expect high conversion efficiency for modest values of $\ncav$. Specifically, operation at $\ncav\approx 13$ and $3$ is expected to yield $\etaOM \approx 50\%$ in devices I and II, respectively. The bandwidth of the transducer is limited by the total acoustic linewidth, which is dominated by the optomechanical backaction $\gamma_\text{m} \approx \gammaOM$ in the regime of high conversion efficiency.

To characterize the transducer-added noise under continuous-wave laser excitation, we use the measurement setup shown schematically in Figure \ref{fig3}(a), where we perform single-photon counting at the optical cavity resonance frequency. The data in Figures \ref{fig3} (b, c) show $\nth$ measured with varying $\ncav$ for devices I and II, respectively. In the high power regime, specifically for $\ncav > 500$, the $\nth$ data has been corrected for the limited bandwidth of the sideband filter when the  acoustic bandwidth $\gamma_\text{m}$ starts to approach the sideband filter linewidth (see supplementary text, section III for details). The top x-axis displays the corresponding $\gammaOM$ for side-coupled devices. The green and purple curves in these panels show results from similar measurements of thermal occupancy of the acoustic mode performed previously on 1D OMCs \cite{maccabe2020nano} and butt-coupled 2D OMCs \cite{ren2020two}. We observe that side-coupled 2D OMCs allow for lower thermal noise across a wide range of input optical powers. A minimum $\nth$ of 0.28 $\pm$ 0.01 is achieved for device I at $\ncav = 2030$, corresponding to an optomechanical transduction efficiency of 98.6 $\pm$ 0.2\%, and a bandwidth of $\gammaOM \approx 6$ MHz.

We model the thermal occupancy $\nth$ using the heating model introduced in the previous section. However, in contrast with the  measurements with $\Delta=0$ in the previous section, the majority of the incident power is reflected under detuned operation with $\Delta=-\Omega_\text{m}$, and parasitic optical absorption in the coupling waveguide can contribute substantially to the hot bath. To include this effect in our heating model, we define an effective photon occupation associated with coupling waveguide, $\nwg$, varying linearly with the input power as \(\nwg = \beta \Pin\) for some fixed constant, $\beta$ and the input power, $\Pin$. We then add the contributions from both cavity and waveguide components and define the parameters of the modified hot bath as $\nbathp[\ncav, \beta] \rightarrow \nbathp[\ncav+\beta\Pin]$, and similarly for $\gammap$. The thermal occupation of the acoustic mode is then given by
\begin{equation}
    \nth=\frac{\gamma_\text{p}[n_\text{c},\beta] n_\text{p}[n_\text{c},\beta]+\gamma_0 n_\text{f}}{\gamma_\text{p}[n_\text{c},\beta]+\gamma_0+\gammaOM[n_\text{c}]}
    \label{eq:nth_theory}
\end{equation}

The solid blue and orange curves in Figures \ref{fig3}(b) and (c) show the predicted thermal phonon occupancy if the waveguide-heating contribution, $\beta$ were set to zero. We see that the experimental data from both devices I and II is in reasonable agreement with the $\beta= 0$ curves. This indicates that waveguide-related heating is negligible in side-coupled 2D-OMC geometry. In comparison, for the butt-coupled 2D OMCs, $\beta$ was measured to be $\SI{15}{\micro\watt}^{-1}$ \cite{ren2020two}. 

While $\nth$ in Equation \ref{eq:nth_theory} depends on the optomechanical device parameters $\gamma_0$ and $\gammaOM$, improvements purely based on geometric modifications of the device platform under detuned operation $\Delta = -\Omega_\text{m}$ can be studied using the heating rate, $\gammap \nbathp/2\pi[\ncav, \beta]$, and is plotted in Figure \ref{fig3}(d) for various devices.
For 1D-OMC, the heating rate scales linearly ($\gammap \nbathp \propto \ncav$), similar to the cooling rate ($\gammaOM \propto \ncav$), which results in the saturation of $\nth$ to a few phonon levels as $\ncav$ increases. 2D geometries are expected to have lower magnitude of the heating rate due to larger thermal contact with the cold bath. However, due to a large $\beta = \SI{15}{\micro\watt}^{-1}$ , the butt-coupling design exhibits almost identical heating performance ($\gammap \nbathp \propto \ncav^{0.9}$) to that of 1D-OMC for $\ncav < 300$, {and only becomes sublinear for $\ncav>300$ ($\gammap \nbathp \propto \ncav^{0.6}$)}. In contrast, for side-coupled geometries, the heating rate scales sub-linearly ($\gammap \nbathp \propto \ncav^{0.6}$) for all $\ncav$ range measured and has a lower magnitude. Specifically, for device II, the heating rate is reduced by approximately six-fold compared to the butt-coupling design for $\ncav > 300$.




\noindent 
\begin{figure}[tp!] 
\begin{center}
\includegraphics[width=\columnwidth]{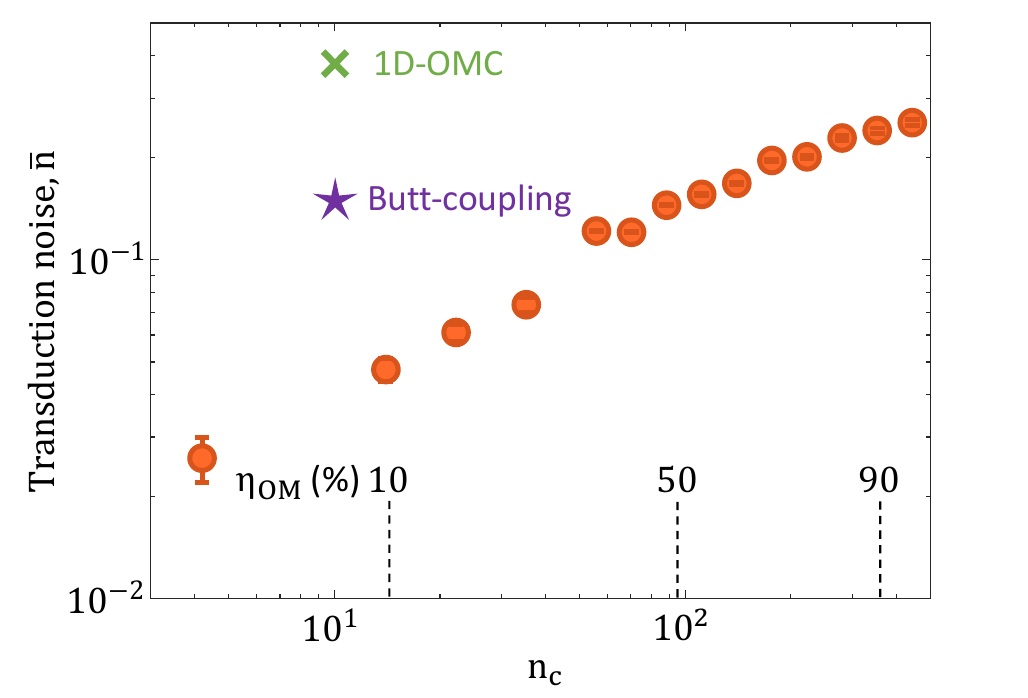}
\caption{\textbf{Phonon-to-photon transduction under pulsed laser excitation}. Converter-added noise, $\Bar{n}$ (red data points) as a function of the peak intra-cavity pump photon number, $\ncav$. All noise measurements are performed on device II with rectangular optical pump pulses with a pulse width of 500 ns at a repetition rate of 250 Hz. For comparison, $\Bar{n}$ is shown for Butt-coupling design \cite{ren2020two}, and 1D-OMC \cite{maccabe2020nano}. Dashed lines indicate the $\ncav$ value for  transduction efficiency $\etaOM$ = 10\%, 50\%, and 90\% for device II. For $\ncav=1$, on chip input power is $\Pin=\SI{0.28}{\micro\watt}$.}

\label{fig4}
\end{center}
\end{figure}

\noindent\textbf{Phonon-to-photon transduction under pulsed laser excitation.} Pulsed transduction schemes are often preferred due to the delayed heating response of the acoustic resonator \cite{meenehan2015pulsed}, which allows for the initialization of the optomechanical transduction pulse prior to the onset of heating. {We characterize the performance of side coupled 2D-OMC as a transducer in pulsed mode by sending short rectangular pulses on red sideband. A pulse duration of $\tau_d=$ 500 ns is selected to account for the finite rise time of the sideband filters ($\approx$ 200 ns).} The transduction efficiency in the pulsed mode is given by
\begin{equation}
    \etaOM=\frac{\gammaOM}{\gammaOM+\gamma_0+\gamma_\text{p}}\left(1-e^{-(\gamma_0+\gamma_\text{p}+\gammaOM)\tau_\text{d}}\right),
    \label{eq:etaOM}
\end{equation}
 Figure \ref{fig4} shows the transduction noise $\Bar{n}$ for device II in pulsed scheme along with similar measurements performed previously on 1D-OMC \cite{maccabe2020nano} and butt-coupling design \cite{ren2020two} at $\ncav$ = 10. The $\ncav$ value corresponding to the optomechanical conversion efficiency $\etaOM$ = 10\%, 50\%, and 90\% for device II are indicated with dashed lines. Given the low intrinsic linewidth of the acoustic resonator for this device ($\gamma_0/2\pi$= 0.97 kHz), we used a repetition rate of 250 Hz to allow the acoustic mode to sufficiently thermalize to the cold bath between successive optical pulses. For the highest optical power we could send to the device ($\ncav$ = 444), we measured $\Bar{n}$ = 0.25 $\pm$ 0.01 which corresponds to a transduction efficiency of $\etaOM$ = 93.1 $\pm$ 0.8\%. This constitutes a significant enhancement across transduction metrics over 1D-OMC with ($\ncav,\Bar{n},\etaOM)\simeq (10,0.4,4\%$) \cite{maccabe2020nano}.

\vspace{2mm}

\noindent\textbf{Discussion}



The side-coupled 2D-OMC design presented here reduces optical absorption-induced thermal noise in OMCs, a major obstacle in quantum application requiring operation at millikelvin temperatures. We anticipate that further reduction in thermal noise can be achieved through surface passivation techniques \cite{borselli2006measuring}, potentially minimizing optical absorption via surface defect reduction. 

This result unlocks new possibilities for more quantum-coherent applications using OMCs. The compact form factor and long-lived acoustic mode of such devices are attractive for applications such as quantum memory for telecom photons \cite{wallucks2020quantum}. {While we achieved a low intrinsic decay rate of 0.97 kHz for the acoustic resonator, we anticipate that embedding the 2D-OMC in a cross phononic shield may lead to drastically lower intrinsic decay rates \cite{maccabe2020nano,ren2020two}.} 
The low thermal occupation of the acoustic resonator presents an opportunity to investigate two-level systems (TLS) in amorphous solids \cite{maccabe2020nano,chen2023phonon,cleland2023studying}, further expanding the potential of OMCs in quantum technologies.

The improved performance of our side-coupled 2D-OMC design in pulsed operation promises significant advancements in single-photon heralding and remote entanglement generation. For a realistic pulse repetition rate of 10 kHz and a total detection efficiency of 5\%, we estimate a single-photon heralding rate of 465 Hz, a substantial increase compared to the 20 Hz achievable with 1D-OMCs. Furthermore, in a two-node remote entanglement experiment, we project a photon coincidence rate of 21 Hz, representing a $\sim$500-fold enhancement over the 0.04 Hz rate achievable with 1D-OMCs.

In the context of microwave-to-optical quantum transduction, side-coupled 2D-OMC-based piezo-optomechanical transducers offer the potential for large-bandwidth, low-noise, and near-unity efficiency conversion between microwave and optical signals. While this study has focused on improved thermal handling in silicon-based devices, piezo-optomechanical transducers utilize heterogeneously integrated platforms with optically robust microwave resonators \cite{meesala2024non,jiang2023optically,weaver2023integrated}, potentially involving distinct heating mechanisms. Nevertheless, our findings establish a performance benchmark for microwave-to-optical transduction. By connecting such a transducer to an off-chip qubit module \cite{delaney2022superconducting,van2023high,arnold2023all}, high optical powers can be employed without compromising superconducting qubit coherence. Additionally, techniques like atomic force microscope (AFM) nano-oxidation can be used to precisely match frequencies between remote piezo-optomechanical systems \cite{hatipoglu2024situ}, paving the way for optically mediated remote entanglement of superconducting qubit nodes.\\


\bibliographystyle{IEEEtran}
\bibliography{arxiv_version}


\vspace{2mm}
\noindent\textbf{Acknowledgements}\\ 
The authors thank Matthew Shaw and Boris Korzh for providing single photon detectors in this work. We thank Piero Chiappina for helpful discussions. This work was supported by the ARO/LPS Cross Quantum Technology Systems program (grant W911NF-18-1-0103), the U.S. Department of Energy Office of Science National Quantum Information Science Research Centers (Q-NEXT, award DE-AC02-06CH11357), the Institute for Quantum Information and Matter (IQIM), an NSF Physics Frontiers Center (grant PHY-1125565) with support from the Gordon and Betty Moore Foundation, the Kavli Nanoscience Institute at Caltech, and the AWS Center for Quantum Computing. S.M. acknowledges support from the IQIM Postdoctoral Fellowship.\\
\textbf{Supplementary information} is available in the online version of the paper.  \\
\textbf{Competing interests.} O.P. is currently employed by Amazon Web Services (AWS) as Director
of their quantum hardware program. AWS provided partial funding
support for this work through a sponsored research grant.\\
\textbf{Materials \& Correspondence.} Correspondence and requests for materials should be sent to OP (opainter@caltech.edu).

\clearpage

\renewcommand{\thefigure}{S\arabic{figure}}
\setcounter{figure}{0} 

\onecolumngrid


{
\centering
\LARGE

\textbf{Supplementary Information}\\[1cm] 
}
\section{Device Parameters}
\label{tab:device_comparison}

\begin{table}[h!]
\centering
\begin{tabular}{|c |c| c|}
\hline
Parameter & Device I & Device II \\
\hline
$\lambda_0$ (nm)  & 1553.4 & 1568.3 \\
$\kappa_\text{i}/2\pi $ (MHz)  & 706 & 595 \\
$\kappa_\text{e}/2\pi$ (MHz)& 444 & 295 \\
$\Omega_\text{m}/2\pi$ (GHz) & 10.44 & 10.29 \\
$\gamma_0/2\pi$ (kHz)  & 21.46 & 0.97 \\
$g_{\text{OM}}/2\pi$ (kHz)  & 919 & 742 \\
\hline
\end{tabular}


\caption{\textbf{Summary of measured device parameters:} $\lambda_0$ is the optical mode wavelength, $\kappa_\text{i}$ is the intrinsic linewidth of the optical cavity, $\kappa_\text{e}$ is the coupling rate between the optical cavity and the optical coupling waveguide, $\Omega_\text{m}$ is the acoustic frequency, $\gamma_0$ is the intrinsic linewidth of acoustic resonator measured using ringdown technique at 10mK temperature \cite{maccabe2020nano}, $\gOM$ is the vacuum optomechanical coupling rate.}
\label{tab:device_comparison}
\end{table}

\section{Mode occupancy calibration}

The cavity photon occupation $\ncav$ is calibrated using the input power at the device $\Pin$ with the following relation:

\begin{equation}
    \ncav=\frac{\Pin}{\hbar\omega_\text{l}}\frac{\kappa_\text{e}}{\Delta^2+\kappa_\text{t}^2/4},
\end{equation}
where $\omega_\text{l}$ is the pump frequency, $\kappa_\text{e}$ is the coupling rate between optical cavity and the optical coupling waveguide, $\kappa_\text{t}=\kappa_\text{i}+\kappa_\text{e}$ is the measured total linewidth of the optical cavity, and $\Delta=\omega_\text{l}-\omega_\text{c}$ is the detuning of the laser from the cavity frequency $\omega_\text{c}$. To calibrate the acoustic mode phonon occupancy to the measured photon count rate on the single photon detector, we perform sideband asymmetry measurements \cite{meenehan2015pulsed}. The measured count rate for red and blue detuned pump laser are given by

\begin{equation}
    \Gamma(\Delta = \pm \Omega_\text{m}) = \gammaDark + \gammaPump + \gammaSB\left(\nth + \frac{1}{2}(1 \pm 1)\right),
\end{equation}
where $\nth$ is the occupation of the acoustic resonator. $\gammaDark$ is the dark count rate of the SPD, $\gammaPump$ is the pump bleedthrough, and $\gammaSB$ is the detected photon scattering rate per phonon. $\gammaSB$ relates to the optomechanical scattering rate $\gammaOM=4\gOM^2\ncav/\kappa_\text{t}$ through the external detection efficiency, $\gammaSB=\eta_{\text{ext}}\gammaOM$. We use pulsed laser with repetition rate much smaller than the intrinsic decay rate ($R \ll \gamma_0$) such that the $\nth \ll 1$ at the beginning of each pulse. $\gammaSB$ can then be extracted by taking the difference between the count rate at the beginning of the pulse. For pulsed transduction measurements with short optical pulses shown in Figure 4 of the main text, the count rates for the red- and blue-detuned pump laser are averaged over the duration of the optical pulse. This averaging accounts for any finite rise time in the sideband scattering rates.

The hot bath measurements in the main text are performed with laser on cavity resonance $\Delta=0$. The count rate $\Gamma(\Delta = 0)$ relates to $\gammaSB$ through the sideband resolution factor $(2\Omega_\text{m}/\kappa_\text{t})^2$

\begin{equation}
    \Gamma(\Delta = 0) = \gammaDark +  \gammaPump + \left(\frac{\kappa_\text{t}}{2\Omega_\text{m}}\right)^2\gammaSB n_\text{m} ,
\end{equation}
where $n_\text{m}$ is the measured thermal occupation in the absence of the optomechanical backaction as defined by equation 1 of the main text.


\section{Device Fabrication and Optimization} 
\label{App:device_fabrication}

The devices in this study were fabricated on a silicon-on-insulator (SOI) platform with a silicon device layer thickness of 220 nm and the oxide layer thickness of 3 µm. The device geometry was defined using electron-beam lithography followed by Pseudo-Bosch dry-etch process to transfer the pattern through the 220 nm Si device layer. Photoresist (SPR-220) was then used to define a 100 µm deep ‘trench’ region into the handle silicon of the chip for coupling light to the on-chip waveguides using a lensed fiber. The devices were then undercut using a vapor-HF etch and cleaned in a piranha solution (3:1 H2SO4:H2O2) before a final vapor-HF etch to remove any native and chemically-grown oxide.

\begin{figure}[h!]
    \centering
    \includegraphics[width=\textwidth]{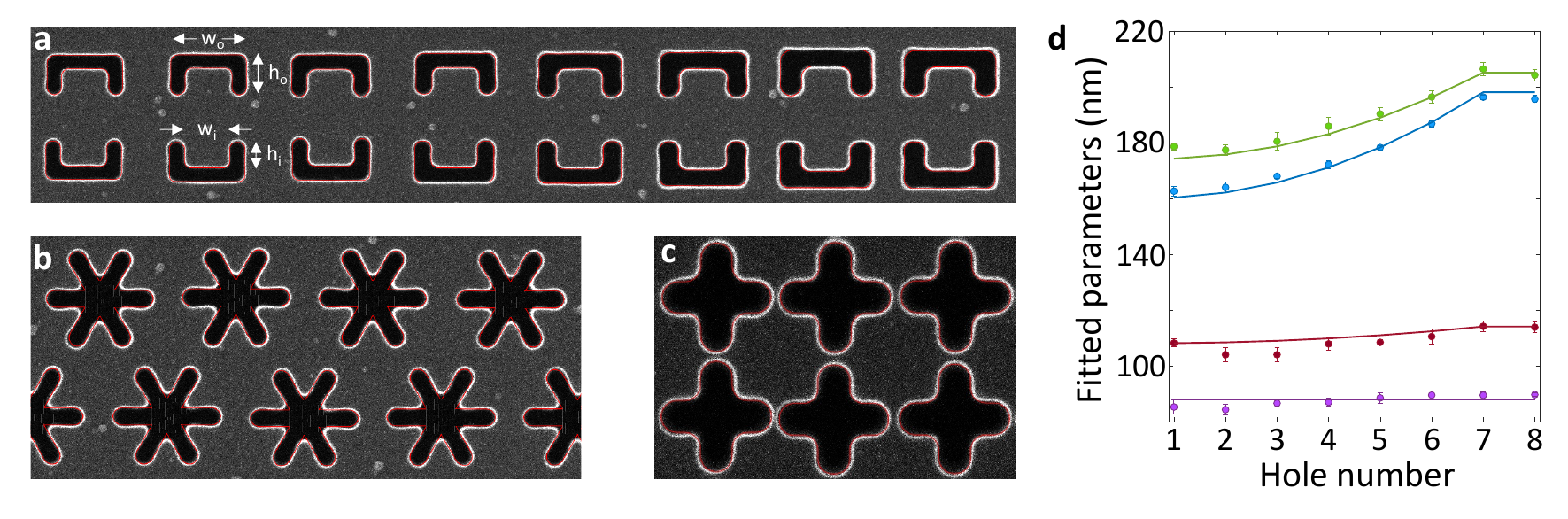}
    \caption{\textbf{Device optimization using image processing and feedback. a,} Helium Ion Microscope (HIM) image of the defect region of the OMC cavity. The fitted features in red are overlaid on top. \textbf{b, c} fitted HIM image of the snowflake and cross shield region. \textbf{d,} Plot showing various measured parameters (circles) from (a) and the corresponding design parameters (solid lines).}
    \label{fig:SI_image_processing}
\end{figure}

The device geometry optimization is aided by image processing. Dimensions of various features are measured and fed back to the subsequent fabrication runs. Figure \ref{fig:SI_image_processing} provides an example of this process using the Helium Ion Microscope (HIM) images of the fabricated geometry. Parameters $h_i$, $w_i$, $h_o$ and $w_o$ are fitted to the designed optomechanical defect parabola. The finite radius of curvature for all features are extracted and fed back into the Finite Element Method (FEM) simulations to optimize the optical and acoustic bandstructure.

\section{Measurement setup}
\label{App:measurement_setup}

\begin{figure*}[tp!]  
\begin{center}
\includegraphics[width=\columnwidth]{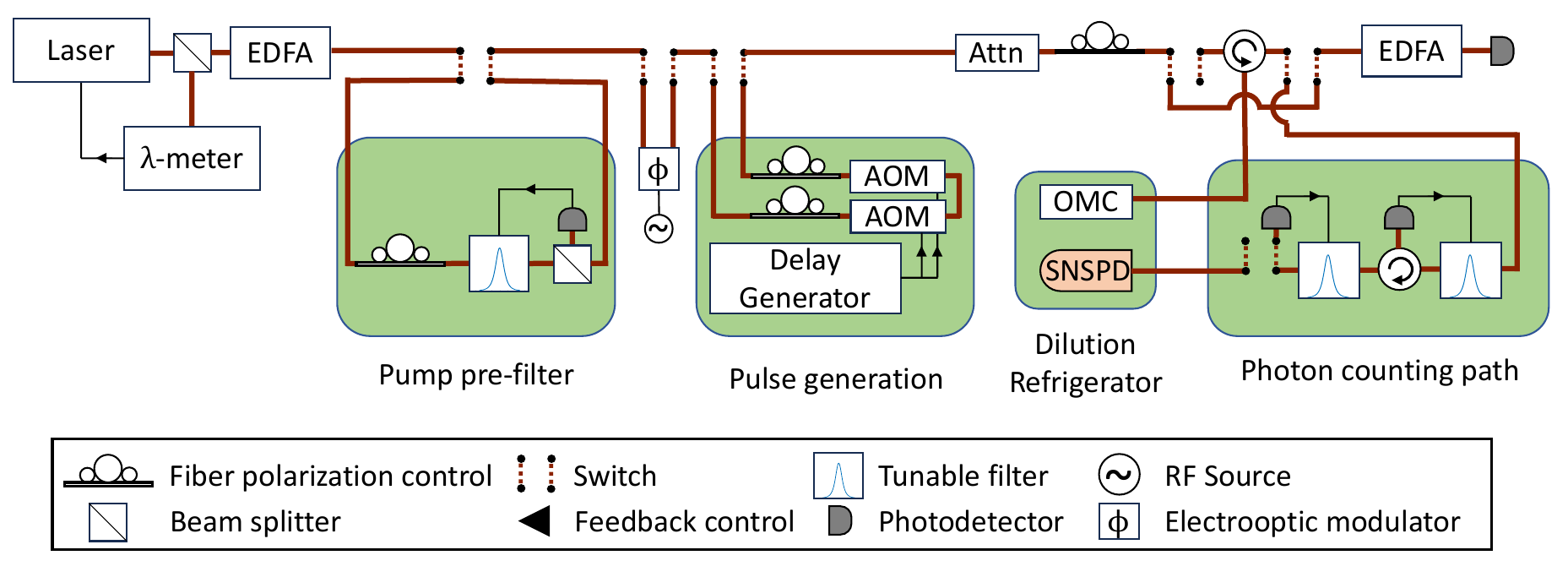}
\caption{\textbf{Experimental setup.} An external cavity diode laser is used to generate the optical pump signal. A small percentage of laser power is used for wavelength locking using a wavemeter ($\lambda$-meter). The laser is then passed through an Erbium Doped Fiber Amplifier (EDFA) to amplify the power, followed by a $50$~MHz-bandwidth filter. It then can be switched through an Electro-optic phase modulator ($\phi$) which is used during the locking sequence of the filter-bank in the photon counting path. The laser can then be switched between two different paths: (i) Continuous wave measurements (ii) a pulsed measurements. In the pulsed measurements path, two acousto-optic modulators (AOM) in series are used for generating high-extinction optical pulses which are triggered by a delay generator. Before sending to the device in the dilution refrigerator (DF), the laser is passed through a variable optical attenuator (Attn) to adjust the optical power. Upon reflection, a circulator routes the reflected laser light to either: (i) spectroscopy photo-detectors which are used for measurements of optical and acoustic spectrum, or (ii) a photon counting path consisting of two cascaded Fabry-Perot filters and a SPD operated at $800$~mK.} 
\label{SI_setupFig}
\end{center}
\end{figure*}

The measurement setup used for the cryogenic measurements is shown in Figure~\ref{SI_setupFig}. A fiber-coupled, wavelength-tunable external cavity diode laser is used as the pump for all measurements. A small percentage of the laser output is sent to a wavemeter ($\lambda$-meter) for wavelength stabilization.
The laser is then amplified with an EDFA and then passed through a $50$~MHz-bandwidth tunable fiber Fabry-Perot filter (Micron Optics FFP-TF2) to suppress broadband spontaneous emission noise. An electro-optic phase modulator $(\phi)$ in the bypass path is used to generate a sideband on the laser during the locking sequence for the sideband filters on the detection path. The pre-filtered pump laser could be switched between (i) a continuous-wave measurement path, and (ii) a pulsed measurement path. On the latter path, we used two acousto-optic modulators (AOMs, G\&H Photonics) in series to create optical pulses with extinction $>$120~dB on the pump laser. A digital delay generator (Stanford Research Systems DG645) is used to synchronize AOM drive pulses and trigger signals for the time-correlated single photon counting module (TCSPC, Swabian Time Tagger X). 

Pump laser light is then routed via a variable optical attenuator (Attn) and a circulator into the dilution refrigerator where the sample test assembly is mounted to the mixing chamber plate cooled down to a base temperature of 10mK. A lensed optical fiber is used to couple light into the device under test. The fiber is mounted to a three-axis nanopositioner stack (Attocube Systems) which allows for in situ alignment with respect to the optical coupler on the device. The reflected pump and signal from the device can be directed via the circulator to (i) a room-temperature photodetection path with a slow detector and a fast detector (New Focus 1554B) for optical and mechanical spectroscopy, or (ii) the single-photon counting path. On the single-photon counting path, the light is passed through two tunable, fiber-coupled Fabry-Perot cavities (Stable Laser Systems) to suppress pump light reflected from the device. The filters have a bandwidth of $3.6$~MHz, a free-spectral range of $15$~GHz, and are connected in series with a fiber-optic circulator in between them. This setup provides approximately $100$~dB of extinction for the pump light detuned by $\sim$10~GHz from the resonance frequency.

During measurements, a locking routine is periodically used to verify that the filters are on resonance with the signal frequency. During the locking routine, we bypass the device path and generate sidebands on the pump laser by driving the EOM ($\phi$) with a microwave tone at the mechanical resonance frequency of the device under test. To re-lock the filters, a sinusoidal voltage was used to dither each filter while monitoring its transmission on a slow detector. The DC offsets of the dithering signal are then changed while reducing the voltage amplitude to align the filters to the expected signal frequency from the device. After re-locking, a new round of measurements can be performed.

For single-photon counting, we used a NbN superconducting nanowire single-photon detector developed by the Jet Propulsion Laboratory (JPL). The SPD is mounted on the still plate of the dilution refrigerator at $\sim 800$~mK. The electrical output of the SPD is amplified by a room temperature amplifier circuit \cite{zhu2020resolving} and read out by a triggered single photon counting module (Swabian Time Tagger X). To minimize the dark counts on the SPD, we filter out long wavelength infrared spectrum noise by creating fiber loops before SPD with a diameter of $1.5$ inches. We observed dark count rates as low as $\sim 4$~(c.p.s.) and a SPD quantum efficiency $\etaSPD \simeq 75\%$ in our setup.\par

\section{Sideband filter chain transmission factor}

\begin{figure}
    \centering
    \includegraphics[width=\textwidth]{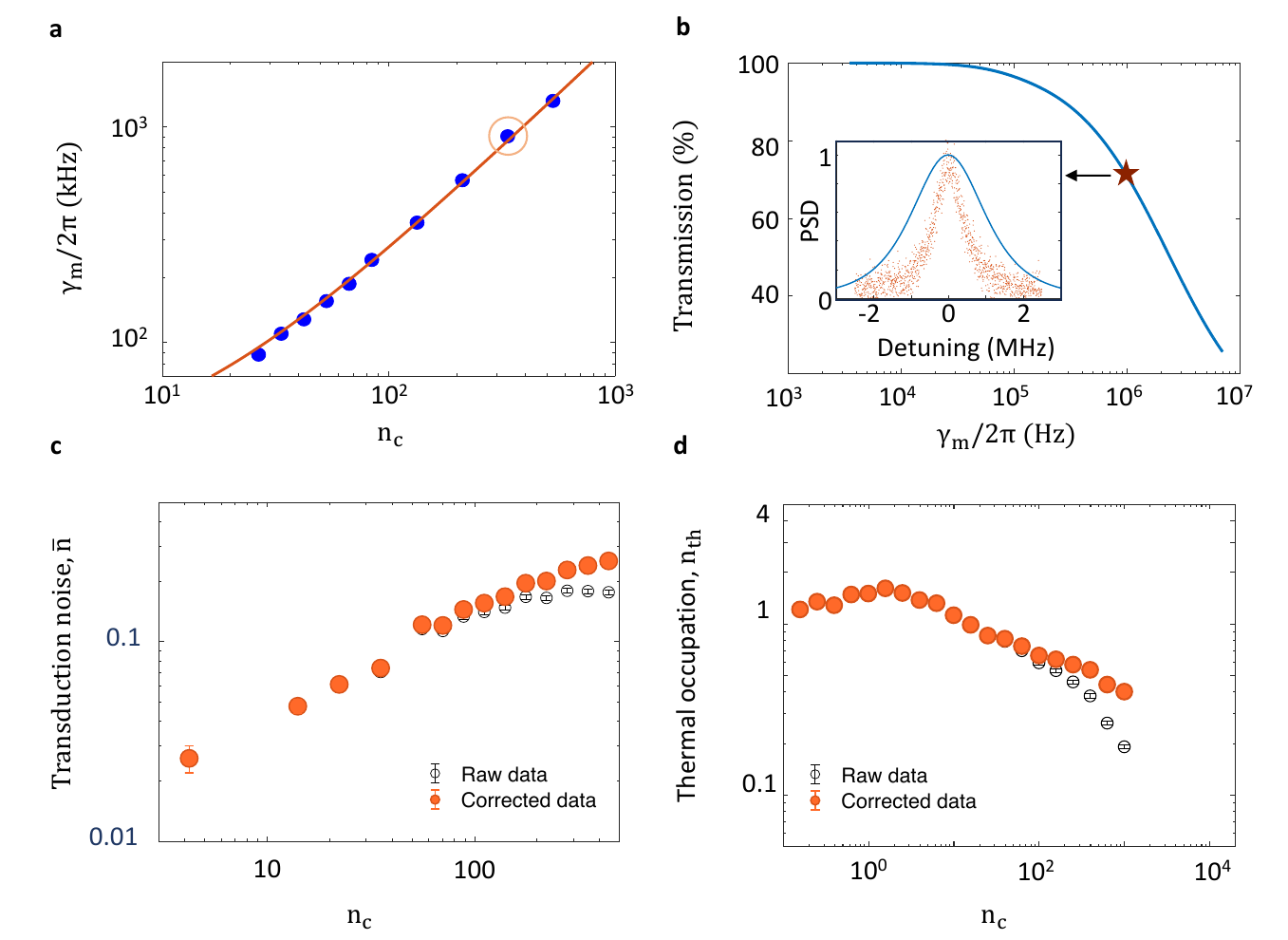}
    \caption{\textbf{Sideband filter chain transmission: a,} total acoustic linewidth as a function of $\ncav$ measured using Electromagnetically induced transparency method (EIT) \cite{safavi2011electromagnetically}. \textbf{b,} transmission through the filter chain as a function of the acoustic linewidth. The inset shows the normalized power spectral density of the filter chain (blue) and that for acoustic sideband (red) for data point marked with circle in panel a ($\ncav=335$). \textbf{c,} phonon occupancy in pulsed mode for device II. The raw data is shown in magenta whereas the corrected data is shown in blue. \textbf{d,} phonon occupancy in continuous wave mode for device II. The black circles show the raw measured data whereas the orange circles show the corrected data.}
    \label{fig:SI_filter_correction}
\end{figure}

The sideband filter chain is parked at the acoustic sideband of the pump to detect the optomechanically scattered photons, while filtering out the pump. However, at large pump powers, the bandwidth of the optomechanical sideband $\gamma_\text{m}$ approaches that of the filter (=3.6 MHz). This overlap leads to a reduced measured count rate, resulting in an inaccurate estimation of phonon occupancy. Figure \ref{fig:SI_filter_correction}(a) shows the acoustic linewidth $\gamma_\text{m}$ as a function of $\ncav$. Measurements are performed using Electromagnetically Induced Transparency (EIT) technique where a pump laser is placed on red sideband and a weak probe scans across a few Megahertz bandwidth across the acoustic sideband at the cavity resonance. $\gamma_\text{m}$ is the sum of the optomechanical backaction $\gammaOM$, intrinsic linewidth $\gamma_i$, coupling rate to the hot bath $\gammap$, and the pure dephasing rate $\gamma_{\phi}$. For two bandpass filters in series, the transmission T of a acoustic sideband with lorentzian profile is given by 

\begin{equation}
    T=\int_{-\infty}^{\infty} H_1(f)H_2(f)\frac{\gamma_\text{m}(f)}{2\pi} df
    \label{eq:T_factor}
\end{equation}

where $H_{1,2}=(1+(\frac{f}{\text{FWHM}_{1,2}/2})^2)^{-1}$ is the normalized spectral response of each filter with $\text{FWHM}_{1,2}$ = 3.6 MHz. The corrected phonon occupancy $\nth$ can then be calculated using measured phonon occupancy $\nth^\star$ using the relation

\begin{equation}
\nth= \nth^\star \times \frac{1}{T}\int_{-\infty}^{\infty} \frac{\gamma_\text{m}}{2\pi} df
\end{equation}

  Figure \ref{fig:SI_filter_correction}(b) shows the transmission factor T as a function of $\gamma_\text{m}/2\pi$. The inset shows the normalized power spectral density of the combined filter chain response (blue), and the acoustic response for $\ncav=335$. Figure \ref{fig:SI_filter_correction}(c) shows the phonon occupancy before (black) and after (orange) correction in pulsed laser mode for device II. Figure \ref{fig:SI_filter_correction}(d) shows the same in continuous wave mode for device II. In the case of $\gammap$ and $\nbathp$ measurements where the pump is placed at the optical cavity resonance, the linewidth of acoustic sideband is much smaller than the filter linewidth ($\gammap+\gamma_0 \ll 3.6$ MHz) for $\ncav$ range measured in this study, so a correction factor is not required.

\begin{figure}[]
\begin{center}
\includegraphics[width=\columnwidth]{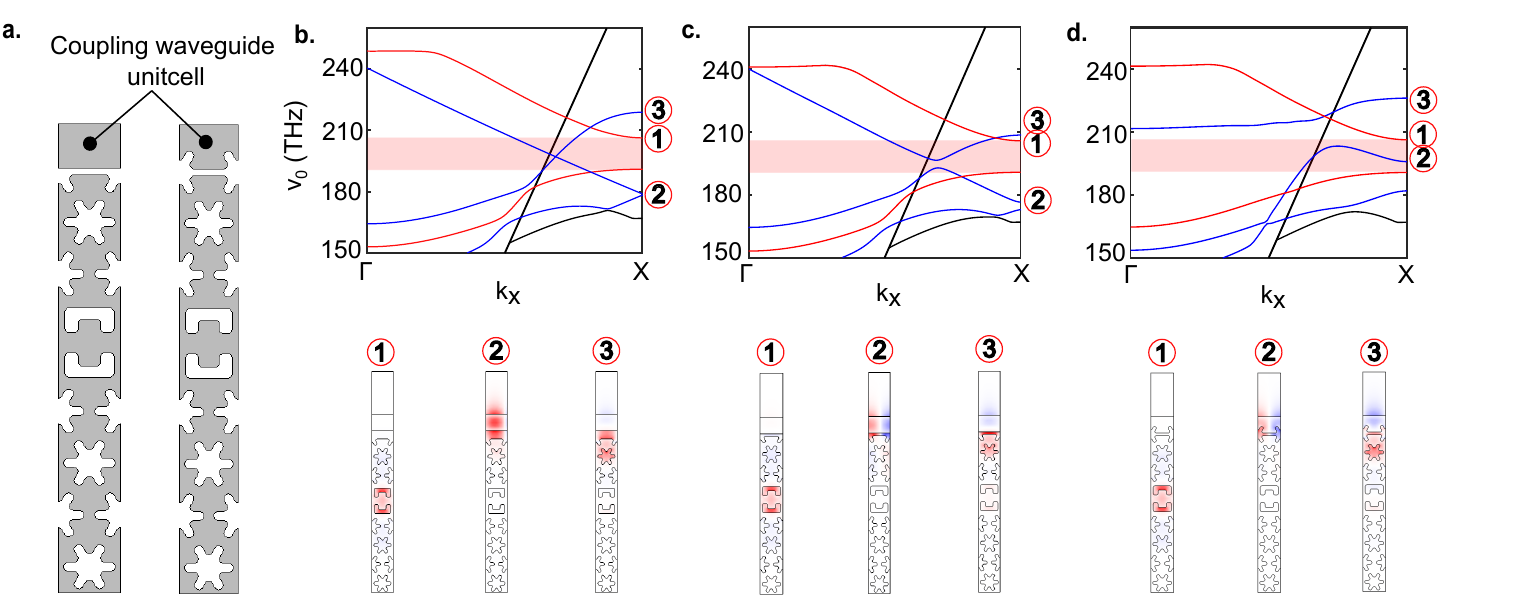}
\caption{\textbf{Optical coupling waveguide design:} \textbf{a}, Schematic shows the supercell design for the 2D-OMC with the coupling waveguide. The left figure shows a conventional waveguide without the snowflake pattern, and the right figure shows the edge-mode coupling waveguide with half snowflakes patterned. \textbf{b, c}, show banstructures for conventional waveguide with 200nm gap, conventional waveguide with 60nm gap , and \textbf{d} edge-mode waveguide with 60nm gap from the 2D structure. The bottom sub figures show the mode structures at X point for the bands labeled with corresponding numbers in the top sub figures.}
\label{SI_coupler_bandstructure_fig}
\end{center}
\end{figure}

\section{Band engineering for low loss optical waveguide}
\label{App:band_engineering}
 As highlighted in the main text, the optical coupling waveguide can significantly influence the hot bath and heat up the acoustic mode as observed in \cite{ren2020two}. Therefore, it is crucial to engineer a mechanically isolated optical coupling waveguide. {Various `mode-converter' designs have been explored in the literature \cite{dutta2016coupling} which convert a waveguide mode into a slot mode. However, since these slot modes have finite overlap with the bulk, further studies are needed to understand isolation of the acoustic resonator from waveguide induced heating. } Here we develop a mechanically detached side-coupled geometry where the waveguide mode of interest is localized primarily in the waveguide slab. We start by cutting the 2D-OMC cavity at the third snowflake row from the center and placing a waveguide slab nearby. Figure \ref{SI_coupler_bandstructure_fig}(a, left) shows the supercell of 2D-OMC with a slab coupling waveguide. The bandstructure combined bandstructure when the waveguide-cavity gap is 200 nm is shown in Figure \ref{SI_coupler_bandstructure_fig}(b). The shaded region shows the relevant optical bandgap of the cavity. The guided modes of the cavity are shown in red whereas the waveguide modes are shown in blue. We identify three relevant modes for the study. Mode 1, 2 and 3 with the corresponding X-point electric field profile are shown underneath the bandstructure. Mode 1 represents the cavity mode of interest, mode 2 represents the waveguide mode of interest, and mode 3 represents the parasitic edge mode localized at the edge of the cut. In this configuration, mode 3 acts as a parasitic mode that reduces the internal quality factor as it crosses the bandgap. Due to large waveguide-cavity separation, waveguide-coupling $\kappa_e$ for mode 2 is very small.
 

Figure \ref{SI_coupler_bandstructure_fig}(c) shows the bandstructure for OMC cavity and conventional waveguide for waveguide-cavity gap of 60 nm. In this configuration, mode 2 weakly hybridizes with mode 3 due to close proximity, however mode 3 still acts as a parasitic mode since it crosses the cavity frequency. This configuration provides large $\kappae$, however internal Q factor of the cavity is greatly reduced due to the parasitic mode 3. To move the parasitic edge mode out of the bandgap, we pattern the waveguide with half snowflake pattern as shown in Figure \ref{SI_coupler_bandstructure_fig}(a,right). As shown in Figure \ref{SI_coupler_bandstructure_fig}(d), mode 3 is moved out of the bandgap while mode 2 still remains in the bandgap. This configuration provides large $\kappa_\text{e}$ while removing parasitic modes from the bandgap.

\begin{figure}[]
\begin{center}
\includegraphics[width=\columnwidth]{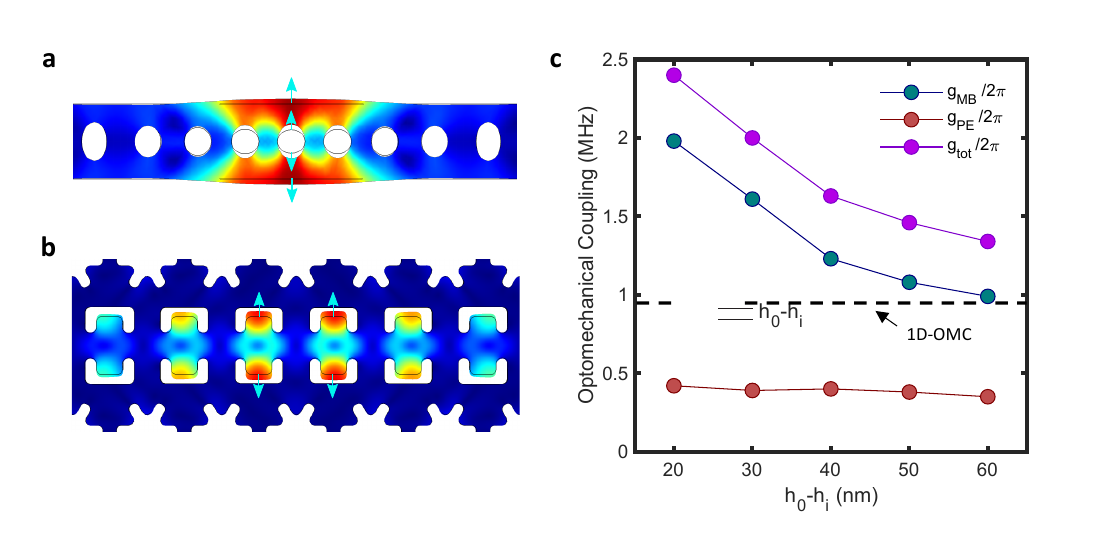}
\caption{\textbf{Theoretical limits on the optomechanical coupling rate} \textbf{a}, 1D OMC breathing mode. The inner moving boundaries of the ellipses have opposite phase compared to the outer boundaries at the edge of the nanobeam. Moving boundary contribution to optomechanical coupling for inner boundaries, $g_{\text{MB in}}/2\pi=341$ kHz whereas that for outer boundaries, $g_{\text{MB out}}/2\pi=-388$ kHz. \textbf{b}, 2D OMC breathing mode is designed to have only outer boundaries that constructively add in phase to the photoelastic optomechanical coupling. \textbf{c,} shows the various contributions to the optomechanical coupling rate as a function of the smallest gap size $h_i-h_o$ for 2D-OMC. The red datapoints show contribution from piezoelectric part $g_{\text{PE}}/2\pi$, the green datapoints show contribution from moving boundary $g_{\text{MB}}/2\pi$ and the magenta shows the total coupling rate $g_{\text{tot}}/2\pi$. The dashed line shows the total coupling rate for 1D-OMC. }
\label{SI_gom_vs_gap_fig}
\end{center}
\end{figure}

\section{Theoretical Limits on Optomechanical coupling rate} 
\label{App:gOM_gap}

In 1D-OMCs, the dominant contribution to the optomechanical coupling rate comes from the photoelastic effect, whereas the contribution from the moving boundary is either negative or very small. Figure \ref{SI_gom_vs_gap_fig}(a) illustrates the various boundaries within a 1D-OMC that contribute to the optomechanical coupling. The inner boundaries of the ellipses exhibit a phase opposite to that of the outer boundaries. The simulated 1D-OMC geometry has a total $g_{\text{tot}}=967$ kHz, out of which 920 kHz is the contribution due to photoelastic coupling $g_{\text{PE}}$. The moving boundary contribution from the inner boundaries (adjacent to ellipses) and outer the boundaries (outer edges), $g_{\text{MB,in}}$ and $g_{\text{MB,out}}$ are -341 kHz and 388 kHz respectively, leading to net 47 kHz moving boundary contribution.

For 2D-OMC, all the moving boundaries are designed to be in-phase, resulting in a net positive contribution to the optomechanical coupling, as depicted in Figure \ref{SI_gom_vs_gap_fig}(b). Furthermore, the moving boundary contribution can be significantly enhanced by reducing the gap size $h_o-h_i$. Figure \ref{SI_gom_vs_gap_fig}(c) shows $g_{\text{PE}}$, $g_{\text{MB}}$ and total optomechanical coupling $g_{\text{tot}}$ as a function of $h_o-h_i$. For comparison, the $g_{\text{tot}}$ for 1D-OMC is shown with dashed lines. For each value of $h_o-h_i$, we run Nelder-Mead Simplex algorithm \cite{ren2020two} to optimize the coupling rate. For gap size of 20 nm, $g_{\text{tot}}$ of 2.45 MHz have been obtained. Experimentally, gap sizes as small as 30 nm have been realized in thin film SOI \cite{liu2022optomechanical} using standard nano-fabrication techniques. Additionally, self assembly techniques have been used to realize gap sizes as small as 2 nm in a bow-tie cavity architecture \cite{babar2023self}. However, we note that with smaller gaps, the optical Q factors become sensitive to surface roughness as more electric field intensity is concentrated in smaller gaps, which will ultimately limit the optomechanical conversion efficiency and bandwidth.

\section{Effect of device parameters on thermal occupation}

\begin{figure}
    \centering
    \includegraphics[width=\textwidth]{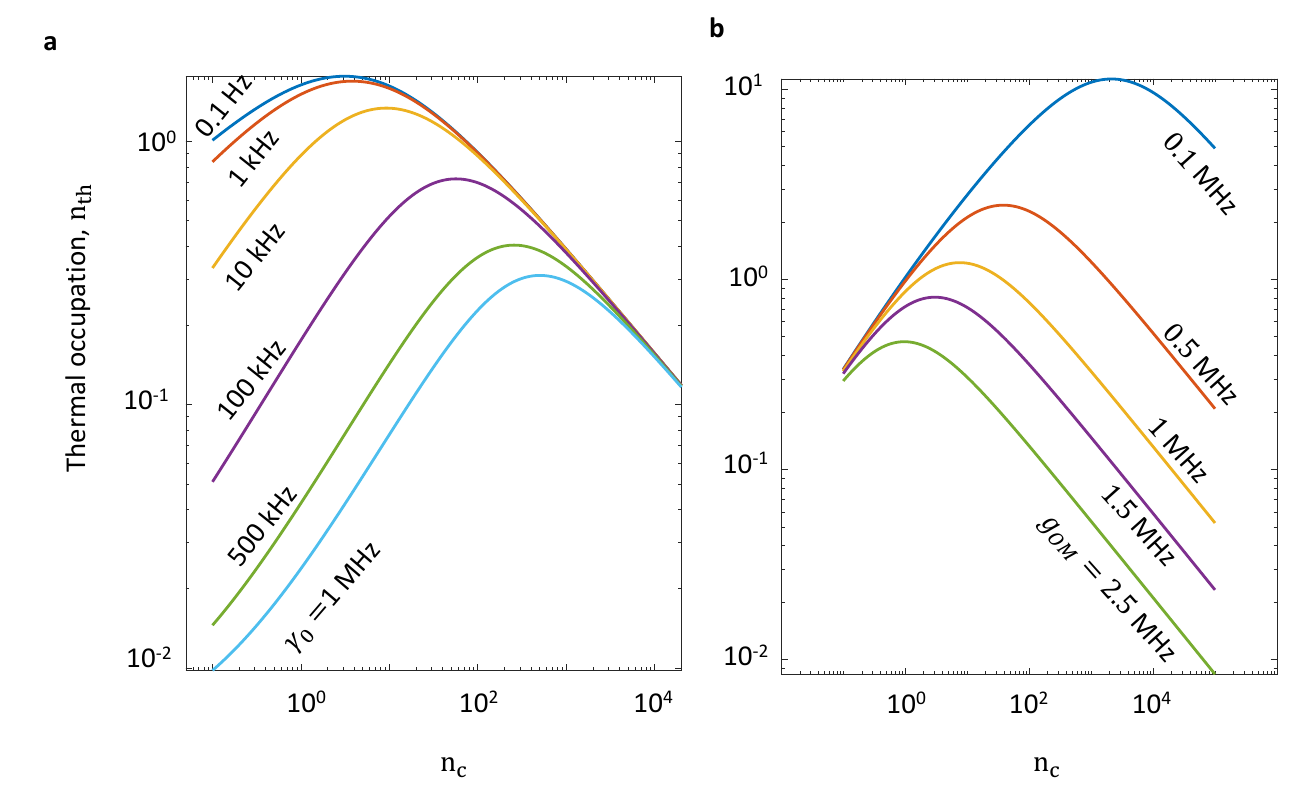}
    \caption{\textbf{Thermal occupation for various device parameters. a,} $n_\text{th}$ as a function of $\ncav$ for various $\gamma_0$. Increasing $\gamma_0$ improves thermalization of the breathing mode to the cold bath (10mK) dictated by equation 2 of the main text, which leads to reduction in $n_\text{th}$. \textbf{b,} $n_\text{th}$ as a function of $\ncav$ for various $\gOM$. The 'turnaround' point shifts to lower $\ncav$ with increasing $\gOM$. For both (a) and (b), device parameters are assumed to be same as that for device I in the main text.}
    \label{fig:nth_gOM_gammaOM}
\end{figure}

In this section, we extend our model for thermal occupation $\nth$ governed by equation 2 of the main text for various device parameters. Figure \ref{fig:nth_gOM_gammaOM}(a) shows $\nth$ vs $\ncav$ plot for various intrinsic linewidth of mechanics $\gamma_0$. With increasing linewidth, the breathing mode thermalizes better with 10 mK bath temperature with occupancy $\nfridge < 10^{-3}$, thereby reducing $\nth$ in the low $\ncav$ regime. Figure \ref{fig:nth_gOM_gammaOM}(b) shows $\nth$ vs $\ncav$ plot for various $\gOM$. In the low power regime ($\ncav<1$), $\nth$ does not show significant difference. However, the 'turnaround' point above which the backaction becomes dominant, shifts to lower $\ncav$ values with increasing $\gOM$. In the high $\ncav$ regime, the $\nth$ shows a significant reduction in amplitude with increasing $\gOM$. The parameters $(\kappa_\text{i},\kappa_\text{e}, \nbathp,\gammap)$ in both panel (a) and (b) of \ref{fig:nth_gOM_gammaOM} are taken from device I.

\section{Numerical simulations for the density of states in 2D-OMC geometry}
\label{app:DOS_simulations}
\begin{figure}
    \centering
    \includegraphics[width=\textwidth]{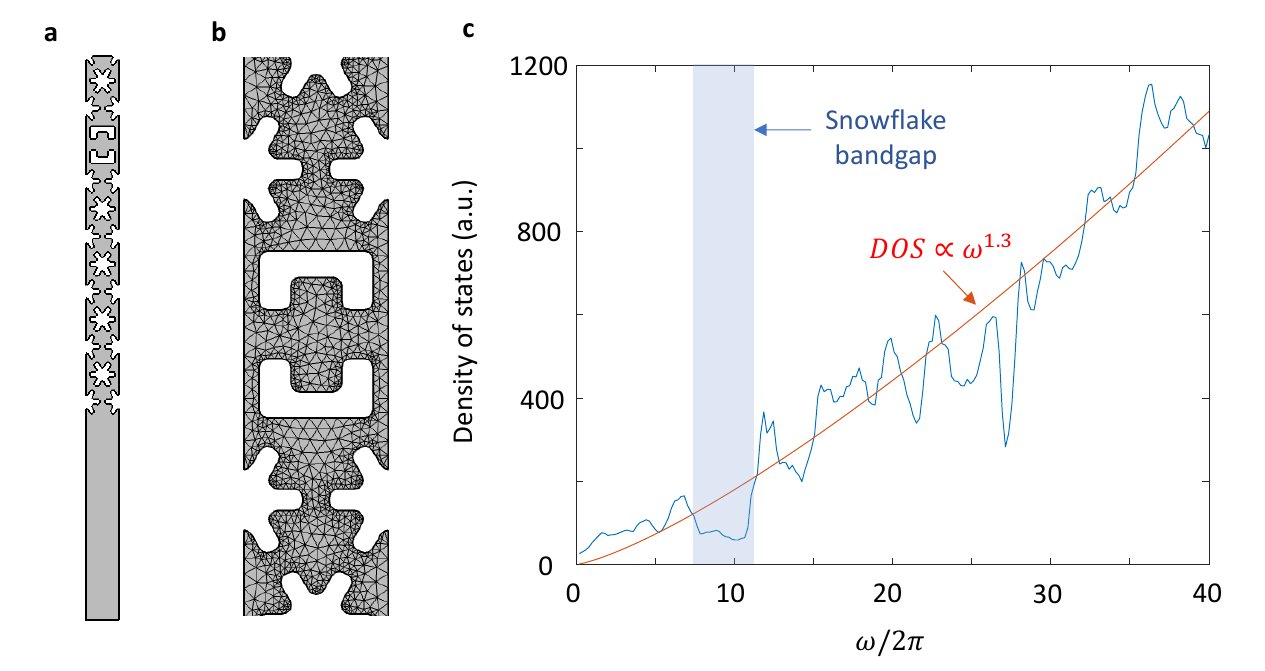}
    \caption{\textbf{Numerical simulations of the acoustic spectrum: a,} unitcell of the 2D-OMC mimicking the exact fabricated  geometry. \textbf{b,} Meshed geometry showing a minimum and maximum element size of 10 nm and 50 nm respectively. \textbf{c,} Density of states as a function of frequency calculated by summing all the acoustic modes from wavevector $\Gamma$ to $X$. The red curve has a power of of 1.3 which corresponds to the dimension of 2.3 for the phononic spectrum. The shaded region shows a dip in DOS due to the bandgap of the snowflakes.}
    \label{fig:DOS_vs_freq}
\end{figure}

To gain insight into the phonon hot bath spectrum and its interaction with the acoustic breathing mode, we employed Finite Element Method (FEM) simulations, targeting all acoustic modes within the structure up to 40 GHz. Given the hot bath temperature range of interest (4 to 10 K), all phonon modes below 1 THz need to be considered. However, the substantial geometry and mesh size requisites pose a computational challenge for simulating frequencies beyond 40 GHz. The FEM simulation setup utilized a supercell geometry, shown in Figure \ref{fig:DOS_vs_freq}(a). The coupling waveguide is excluded from the simulations due to its mechanical separation from the cavity. The configuration includes 2.5 and 9 rows of snowflake patterns above and below the fishbone unit cell, respectively. An additional 3 micron silicon bulk layer beneath the 9 rows accounts for the undercut from the vapor HF release process during device fabrication. Boundary conditions are specified as fixed for the -y edge and free for the +y edge. Figure \ref{fig:DOS_vs_freq}(b) illustrates the geometry prepared for FEM simulations, defined by mesh sizes ranging from a minimum of 10 nm to a maximum of 50 nm. This mesh size ensures adequate resolution across the spectrum, capturing several points per wavelength at the highest frequency analyzed.

Figure \ref{fig:DOS_vs_freq} (c) shows the density of states (DOS) as a function of frequency for the simulated structure, where DOS is calculated by summing all acoustic modes spanning wavevector range $\Gamma$ to $X$. For visual clarity, frequency bin size of 200 MHz is used for plotting. The shaded region shows a reduced DOS due to the bandgap of snowflake structures. The solid line indicates a power law relationship of DOS \(\propto \) $\omega$\(^{1.3}\), suggesting that the phonon bath exhibits effective dimension of 2.3.

\section{Theoretical Model for Hot Bath} 
\label{App:theory_hot_bath}
The local hot bath plays a key role in determining the thermal occupation of the acoustic resonator in the presence of laser light. During the plasma etching of the device, which involves $\text{SF}_6$ and $\text{C}_4\text{F}_8$ chemistry, ion implantation creates defect states at the surface of the silicon. The hot bath is thought to be generated as these defect states, with eV energy levels, undergo phonon assisted decay to create a shower of high frequency phonons. Due to very high density of states, these phonons thermalize among themselves at a rate $\gamma_{\text{mix}}$ much larger than coupling rates with the external environment and hence have a common temperature $T_\text{p}$. The acoustic breathing mode at 10.3 GHz lies among discrete phonon modes and is protected by the phononic bandgap. Despite having spatial overlap between the hot bath and the acoustic breathing mode, the two can be considered to be independent of each other due to small coupling rate rate $\gammap \ll \gamma_{\text{mix}}$.

To derive the expression of $\gammap$, we consider a simple 3-phonon scattering model where the acoustic breathing mode $\Omega_\text{m}$ couples to high frequency modes $\omega_1$ and $\omega_2$ in the phonon bath, such that $\omega_2-\omega_1=\Omega_\text{m}$. Using first order perturbation theory, the scattering rates in and out of the mode of interest are given by $\Gamma_+=A(n_\text{m}+1)n_2 (n_1+1)$ and $\Gamma_-=An_\text{m} n_1(n_2+1)$ respectively, where $n_\text{m}$, $n_1$, and $n_2$ are the number of phonons in the corresponding modes, and A is the anharmonicity matrix of the lattice. For simplicity, we assume A to be frequency independent. The rate of change of $n_\text{m}$ is then given by
\begin{equation}
    \dot{n}_\text{m}=\Gamma_+-\Gamma_-=-\gammap^*(n_\text{m}-\nbathp^*)
    \label{eq:n_m_dot}
\end{equation}
where $\nbathp^*=An_2(n_1+1)/\gammap^*$ is the effective occupancy in the steady state and $\gammap^*=A(n_2-n_1)$ can be thought of as the rate at which the breathing mode reaches occupancy $\nbathp^*$. 
The total rate $\gammap$ would be the sum of individual $\gammap^*$ across the hot bath frequency spectrum.
\begin{equation}    \gammap=\sum_{n_1,n_2}A(n_1-n_2)
\end{equation}

In thin film patterned systems, the density of states is significantly modified in the low frequency regime such that $\omega_2-\omega_1=\Omega_\text{m}$ is not satisfied in the continuum limit. We take the continuum limit for $\gammap$ above a certain cutoff frequency $\omega_\text{c}$ where the density of states is large,

\begin{equation}    \gammap=\int_{\omega_\text{c}}^{\infty}d\omega A\rho[\omega]\rho[\omega+\Omega_\text{m}](n_\text{b}[\omega]-n_\text{b}[\omega+\Omega_\text{m}]),
    \label{eq:gammaP_integral}
\end{equation}
here, $\rho[\omega]$ is the phonon density of states (DOS) at frequency $\omega$ and $n_\text{b}$ is the Bose Einstein occupation factor. Due to the low DOS below $\omega_\text{c}$, the effective ground state frequency for the hot bath is shifted to $\omega_\text{c}$,

\begin{equation}
    n_\text{b}[\omega]=\frac{1}{e^{\hbar(\omega-\omega_\text{c})/k_\text{B}T_\text{p}}-1}
\end{equation}

\begin{figure}
    \centering
    \includegraphics{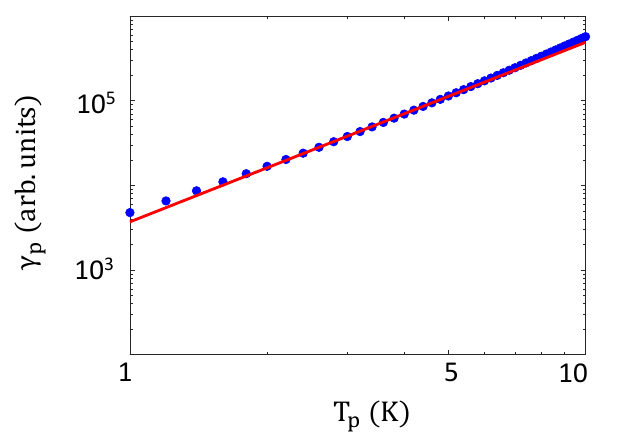}
    \caption{\textbf{Scaling of hot bath coupling rate $\gammap$ as a function of bath temperature.} The blue dots show theoretical value of $\gammap$ obtained by numerically integrating equation \ref{eq:gammaP_integral}. The solid red curve shows a theoretical power law dependency $\gammap \propto T_\text{p}^2$. The dimensionality of phonon bath is taken as 2.3 in accordance with FEM simulations in supplementary section 10, and the cutoff frequency $\omega_\text{c} =$ 20 GHz is chosen.}
    \label{fig:gammaP_vs_temperature_theory}
\end{figure}

For a d-dimensional Debye phonon bath, the DOS follows the power law relation, $\rho[\omega]\propto \omega^{d-1}$. We numerically integrate equation \ref{eq:gammaP_integral} to find $\gammap$ in the temperature range $T_\text{p} \in [1,  10]$ K corresponding to the experimentally observed bath temperature range. Figure \ref{fig:gammaP_vs_temperature_theory} shows the numerically calculated value of $\gammap$ in arbitrary units (blue circles). We use d=2.3 in agreement with the numerical simulations of the acoustic spectrum as discussed in supplementary section \ref{app:DOS_simulations}. We define the cutoff frequency $\omega_\text{c}$ based on the fundamental mode for the thickness of the device layer (t = 220 nm) and the longitudinal velocity ($v_\text{l}$ = 8433 m/s), yielding $\omega_\text{c}/2\pi =v_\text{l}/2t \approx$ 20 GHz. The red curve in Figure\ref{fig:gammaP_vs_temperature_theory} represents the theoretical power law dependency given by $\gammap \propto T_\text{p}^{2}$.

In the large temperature limit $T_\text{p} \gg \hbar\Omega_\text{m}/k_\text{B}$, a generalized power law with d-dimensional DOS is derived in \cite{maccabe2020nano}, and approximately scales as

\begin{equation}
    \gammap \sim\propto T_\text{p}^{2d-2}
\end{equation}

The effective thermal occupancy of the acoustic resonator in the steady state can be calculated by integrating $\nbathp^*$ over the phonon bath frequency spectrum

\begin{equation}
    \nbathp=\frac{1}{\gammap}\int_{\omega_\text{c}}^{\infty} d\omega A\rho[\omega]\rho[\omega+\Omega_\text{m}]n_\text{b}[\omega+\Omega_\text{m}] (n_\text{b}[\omega]+1)
\end{equation}
We use the identity $n_\text{b}[x+x_m](n_\text{b}[x]+1)=(n_\text{b}[x]-n_\text{b}[x+x_m])n_\text{b}[x_m]$ to simplify the above equation.

\begin{align}
     \nbathp &\simeq\frac{n_\text{b}[\Omega_\text{m}+\omega_\text{c}]}{\gammap}\int_{\omega_\text{c}}^{\infty}d\omega A \rho[\omega]\rho[\omega+\Omega_\text{m}](n_\text{b}[\omega]-n_\text{b}[\Omega_\text{m}+\omega_\text{c}])\\
     &= n_\text{b}[\omega+\omega_\text{c}]\\
     &\approx \frac{k_\text{B}T_\text{p}}{\hbar\Omega_\text{m}} \text{  for 
 }  T_\text{p} \gg \hbar\Omega_\text{m}/k_\text{B}   
\end{align}

To relate $\nbathp(T_\text{p})$ and $\gammap(T_\text{p})$ relations above with the $\ncav$ powers laws in Figure 2 of the main text, we utilize Planck's law. In thermal equilibrium, the hot bath acts as a black body with radiated power $P_{\text{out}}$ that scales as $P_{\text{out}}\sim T_\text{p}^{\alpha+1}$, where $\alpha$ is thought to be the effective number of spatial dimensions of the material/structure under consideration. For a linear optical absorption process, we can write the absorbed optical power as a fraction $\eta$ of the optical pump power: $P_{\text{abs}} = \eta P_{\text{in}} = \eta^\prime \ncav $. In thermal equilibrium, the power output from the phonon bath is equal to its input, $P_{\text{out}} = P_{\text{abs}} \sim \ncav$ leading to the relation:
\begin{equation}
T_\text{p} \propto \ncav^\frac{1}{\alpha+1}    
\end{equation}

We then have the following relations for $\nbathp(\ncav)$ and $\gammap(\ncav)$:   
\begin{equation}
    \nbathp\propto T_\text{p} \propto \ncav^{\frac{1}{\alpha+1}}
\end{equation}
and

\begin{equation}
    \gammap\propto T_\text{p}^{2d-2}\propto \ncav^{\frac{2d-2}{\alpha+1}}
    \label{eq:gammaP_dimensionality}
\end{equation}
Table \ref{tab:power_laws} shows comparison of power laws for different devices. For device I, butt-coupling, and 1D-OMC, the $\nbathp$ power law is in the range [0.3,0.33] corresponding to $\alpha$ in the range [2, 2.3]. However, device II show significantly different power law of 0.21 corresponding to $\alpha=$ 3.77. The exact mechanism determining $\alpha$ is still under investigation, however such varying power laws have been observed in \cite{zen2014engineering} which suggests that $\alpha$ is not necessarily equal to the dimension of the phonon bath, $d$, in a thin film phononic structures, and can be larger than 3 dimensions.

The $\gammap$ power laws in the low $\ncav$ regime for Butt-coupling and 1D-OMC have been found to be in the range [0.6, 0.66], which correspond to phonon bath dimensionality of $\approx$ 2. For device II, despite having $\gammap$ power law of 0.39 in the low $\ncav$ regime, the phonon bath dimensionality turns out to be $\approx$ 2 using equation \ref{eq:gammaP_dimensionality}. In the high power regime, the $\gammap$ power law for all measured devices is 0.29 which could be due to a combination of 3-phonon-scattering to Akhiezer damping model \cite{liao2018akhiezer}. Further studies are needed to understand $\gammap$ power laws in the high power regime and will be the subject of future work.
\begin{table}[h!]
\centering
\footnotesize
\renewcommand{\arraystretch}{1.5} 
\begin{tabular}{|c|c|c|c|c|}
\hline

Device & \( \nbathp \) power law & \(\alpha\) & \( \gammap \) power law & \(d\) \\
\hline
\hline
Device I & 0.31 & 2.2 & NA  (low \( \ncav \)) &  NA  (low \( \ncav \)) \\
 & & & 0.29 (high \( \ncav \)) & 1.42 (high \( \ncav \)) \\
\hline
Device II & 0.21 & 3.77 & 0.39 (low \( \ncav \)) & 1.93 (low \( \ncav \)) \\
 & & & 0.29 (high \( \ncav \)) & 1.69 (high \( \ncav \)) \\
\hline
Butt-coupling & 0.3 & 2.3 & 0.6 (low \( \ncav \)) & 1.99 (low \( \ncav \)) \\
 & & & 0.29 (high \( \ncav \)) & 1.48 (high \( \ncav \)) \\
\hline
1D-OMC & 0.33 & 2 & 0.66 (low \( \ncav \)) & 2 (low \( \ncav \)) \\
 & & &  NA  (high \( \ncav \)) &  NA  (high \( \ncav \)) \\
\hline
\hline
\end{tabular}
\caption{Comparison of hot bath dimensionality parameters $\alpha$ and $d$ for various devices. Data for Butt-coupling and 1D-OMC devices are adapted from \cite{ren2020two, maccabe2020nano}.}
\label{tab:power_laws}
\end{table}

\section{Phonon-Photon Quantum transduction with 2D-OMC: Figures of merit}

\begin{table}[h!]
\centering
\begin{tabular}{|@{}|c|c|c|@{}}
\hline
Parameter & \textbf{1D-OMC}\cite{maccabe2020nano} & \textbf{2D-OMC (device II)} \\ 
\hline
\( \ncav \) & 10  & 443 \\
\hline
\( \eta_{\text{OM}} \) & 4\%  & 93\% \\
\hline
$\Bar{n}$ & $\sim$0.4  & 0.25 \\
\hline 
peak $\gammaOM/2\pi$ & 13 kHz & 1.1 MHz\\ \hline
Single photon heralding rate & 20 Hz  & 465 Hz \\ 
(R=10 kHz, $\eta_{\text{ext}}=5\%$) & & \\ \hline
Photon coincidence rate & 0.04 Hz & 21 Hz \\
\hline
\end{tabular}
\caption{\textbf{Comparison of figures of merit for pulsed transduction:} $\eta_{\text{OM}}$ is the phonon-to-photon transduction efficiency. $\Bar{n}$ is the measured noise in the transduction process. Single photon heralding rate is estimated for repetition rate R=10 kHz and detection efficiency $\eta_{\text{ext}}=$ 5$\%$. The coincidence rate is the estimated for a two-node remote entanglement scheme.}
\label{tab:pulsed_transduction_comparison}
\end{table}

\begin{table}[h!]
\centering
\begin{tabular}{|c|c|c|@{}|}
\hline
 \textbf{Parameter}& \textbf{1D-OMC} \cite{maccabe2020nano}& \textbf{2D-OMC (device II)} \\ 
\hline
\( \ncav \) & 569 &  641 \\
\( \etaOM \) &  \(\approx 93\%\) & \(\approx 97\%\) \\
$\nth$ & 4.2   & 0.42 \\
$\gammaOM/2\pi$ & 1 MHz  & 1.6 MHz\\ \hline
\end{tabular}
\caption{\textbf{Comparison of figures of merit for continuous-wave transduction}}
\label{tab:CW_transduction_comparison}
\end{table}

In this section, we evaluate the performance improvements in a 2D-OMC based transducer over their 1D-OMC counterpart. Table \ref{tab:pulsed_transduction_comparison} shows a comparative analysis between 1D-OMC (device A in \cite{maccabe2020nano}) and 2D-OMC (device II in this work) for phonon-photon pulsed transduction. For both measurements, the repetition rate is much smaller than the intrinsic decay rate of the acoustic resonator ($R\ll \gamma_0$) which allows for thermalization with the fridge environment between subsequent optical pulses. For a fair comparison of the single photon heralding rate, we assume a common repetition rate R=10 kHz and total detection efficiency $\eta_{\text{ext}}=5\%$ for both 1D-OMC and 2D-OMC. The resulting single photon heralding rate for 1D-OMC is $R\times \eta_{\text{ext}}\times \etaOM$ = 20 Hz with an added noise of $\Bar{n}\sim 0.4$. For 2D-OMC, the single photon heralding rate is 465 Hz with an added noise of 0.25. {The photon coincidence rate for a remote entanglement experiment involves two nodes and will scale as square of the detection efficiency.} For 1D-OMC, this rate is significantly smaller ($R\times \eta_{\text{ext}}^2\times \etaOM^2$=0.04 Hz); whereas the photon coincidence rate for 2D-OMC in this work is 21 Hz, which constitutes a $\sim$500-fold improvement along with lower added noise.

In case of continuous wave transduction, the figures of merit for different devices are shown in table \ref{tab:CW_transduction_comparison}. The data for 1D-OMC is taken for device I in \cite{maccabe2020nano}. The transduction efficiency $\etaOM$ in continuous wave is given by $\gammaOM/(\gammaOM+\gamma_0+\gammap)$ which remains greater than 90$\%$ for $\gammaOM \gg \gamma_0$. The thermal occupation of the acoustic resonator shows an order of magnitude improvement for device II in this work ($\nth=0.42)$, compared to 1D-OMC ($\nth=4.2$) at similar $\ncav$ levels.

\end{document}